\documentclass[%
prl%
,twocolumngrid%
,floats%
]{revtex4}
\begin{document}

\title{$d_{\rm c}=4$ is the upper critical dimension for the Bak-Sneppen
model}

\author{Stefan Boettcher$^{1}$ and Maya Paczuski$^{2,3}$}
%\thanks{}
\affiliation{
$^1$Physics Department, Emory University, Atlanta, Georgia 30322, USA\\
$^2$Nordita, Blegdamsvey 17, 2100 Copenhagen, Denmark \\
$^3$Department of Mathematics, Imperial College, 180 Queen's Gate,
London, U.K. SW7 2BZ. }
\date{\today}

\begin{abstract}
Numerical results are presented indicating $d_{\rm c}=4$ as the upper
critical dimension for the Bak-Sneppen evolution model.  This finding
agrees with previous theoretical arguments, but contradicts a recent
Letter [Phys.~Rev.~Lett.~{\bf 80}, 5746-5749 (1998)] that placed
$d_{\rm c}$ as high as $d=8$. In particular, we find that avalanches
are compact for all dimensions $d\leq4$, and are fractal for
$d>4$. Under those conditions, scaling arguments predict a $d_{\rm
c}=4$, where hyperscaling relations hold for $d \leq 4$.  Other
properties of avalanches, studied for $1\leq d\leq6$, corroborate this
result.  To this end, an improved numerical algorithm is presented
that is based on the equivalent branching process. %\hfil\break\noindent
PACS number(s): 64.60.Ak, 05.40.-a, 05.65.+b.
\end{abstract}
%\pacs{PACS number(s): 64.60.Ak, 05.40.-a, 05.65.+b.}
\maketitle
The Bak-Sneppen evolution model~\cite{B+S} has received considerable
attention as an archetype of self-organized criticality 
\cite{BTW}, which has been put forward as a general mechanism leading to
many non-equilibrium scaling phenomena observed in
Nature~\cite{Bakbook}.  Although the Bak-Sneppen model was originally
developed as an attempt to interpret paleontological data indicating
co-evolutionary activity in biological evolution~\cite{B+S,PNAS}, it
has also been used to interpret power law distributions in quiescent
periods between earth-quakes~\cite{Ito}. Its generic mechanism of
extremal dynamics~\cite{scaling} has even inspired an algorithm to
approximate combinatorial optimization problems~\cite{EOperc}.

In a recent Letter~\cite{highd}, numerical results for several
critical exponents of the Bak-Sneppen model in high dimensions were
presented. On the basis of that study, and certain scaling arguments,
Ref.~\cite{highd} concluded that the upper critical dimension, where
those exponents obtain mean field values, was $d=8$.  Avalanches,
cascading through the system via nearest-neighbor activation
events, form domains that were found to be fractal for $d>2$. These two
findings are surprising: On one hand, avalanches would proceed on a
filamentary domain where surface sites outnumber bulk sites.  On the
other hand, activity would have to return mostly --- by pure chance
--- to sites already within that domain to keep it from growing
linearly with time, as required in the mean-field limit.

The claims for Ref.~\cite{highd} are in stark contrast to earlier
derived scaling relations (see Tab.~1 in Ref.~\cite{scaling}), which
would predict an upper critical dimension of $d=4$. For instance, the
scaling relation for the avalanche cut-off is  $\sigma=1-\tau+d/D$.
Its mean-field value is $\sigma=1/2$~\cite{meanfield,BoPa1}, and the
avalanche distribution exponent attains
$\tau=3/2$~\cite{meanfield,BoPa1}, while Ref.~\cite{BoPa1} explicitly
derived that $D=4$ is the mean field value for the avalanche dimension
exponent, implicating $d=4$ as the upper critical dimension.  But the
scaling relations in~\cite{scaling} were derived under the clearly
stated assumption that avalanches would always be compact for
$d<d_{\rm c}$, and activity within them would proceed homogeneously
over their domain, with many returns to each site. Ref.~\cite{highd}
argues that avalanches are ramified already for $d>2$, allowing for a
$d_{\rm c}>4$.

In this letter we test the claims of Ref.~\cite{highd} by independent
means. To this end, we have developed an alternative algorithm, which
has the added benefit of improving temporal cut-offs by more than two
decades while eliminating finite lattice size effects entirely. As a
result, we find the assumptions underlying the scaling theory
in~\cite{scaling} intact. In particular, our numerical simulation
results show that avalanches in $d\leq4$ are compact, and that scaling
exponents, when asymptotic behavior is extracted, take on mean-field
values for $d=5$  and 6, very much in contradiction to
Ref.~\cite{highd}.  Thus we show that the Bak-Sneppen model possesses
an upper critical dimension, $d_c=4$, below which hyperscaling
relations hold, similar to equilibrium critical phenomena.  In fact,
we have  studied a whole range of different properties  of the
Bak-Sneppen model for $1\leq d\leq6$ ~\cite{bsd.pre}. Here, we  focus
directly on those properties relevant with regard to the upper
critical  dimension.

The Bak-Sneppen model is very easy to state~\cite{B+S}. It consists of
random numbers $\lambda_{\bf r}$ between 0 and 1, occupying sites
${\bf r}$ on a $d$-dimensional lattice. At each update step, $s$, the
smallest random number $\lambda_{min}(s)$ is located.  That site as
well as its $2d$ nearest neighbors each receive new random numbers
drawn independently from a flat distribution on the unit
interval. This update step is repeated at the site with the next
$\lambda_{min}(s+1)$, and so on. The process inevitably evolves toward
a self-organized critical state~\cite{BTW} in which almost all
$\lambda_{\bf r}$ are larger than a critical threshold $\lambda_{\rm
c}$ (see Tab.~\ref{lambdac}), while those $\lambda_{\bf r}$ below are
part of an avalanche of activity. The current minimum
$\lambda_{min}(s)$ is the ``active'' site while those
$\lambda<\lambda_{\rm c}$ are ``unstable'' and potentially active; all
$\lambda>\lambda_{\rm c}$ are ``stable''. These avalanches are
critical and their properties have been described in terms of scaling
relations~\cite{scaling}.

The obvious way to implement the model  is to specify a lattice, place
a  random number $\lambda_{\bf r}$ on each site ${\bf r}$, keep an
efficient list of those numbers, ordered by size, and repeatedly
replace the smallest $\lambda_{min}(s)$ and its neighbors. In this
method, knowledge of $\lambda_{\rm c}$ is {\it a priori} not necessary
to determine the critical properties of the avalanches. This is, in
fact, the algorithm employed  in Ref.~\cite{highd} and some other,
low-dimensional studies. {\it But in higher dimensions feasible length
scales in the pre-defined lattice, with a fixed number of sites
$N=L^d$, exponentially diminish  with dimension.} For instance,
Ref.~\cite{highd} used $N=2^{16}$, giving $L\!=\!16$ in $d\!=\!4$
(only $L\!=\!4$ in $d\!=\!8$), or a maximal spatial separation of
$r_{\rm max}=d L/2=32$, and a temporal cut-off of maximally $s_{\rm
co}\sim r_{\rm max}^D\approx 10^6$, since $D\leq 4$.

A more efficient way to describe an avalanche utilizes a
generalized branching process~\cite{PMB,scaling} that has been shown
to be exactly equivalent to the Bak-Sneppen model.  In the update
procedure, the values of the
 new random numbers that enter the  system are unrelated
to each other, their predecessors, or any other number  in the
system. Thus, any number that has no chance of becoming active 
in itself (i. e. those with $\lambda>\lambda_{\rm c}$), will not affect
the  dynamics of the critical avalanche through its particular
value. Now, assume  that we are at the beginning of a critical
avalanche, i. e. all numbers in the system are above $\lambda_{\rm
c}$. All we need to know to describe the ensuing critical avalanche is
the existence of a smallest number in the system to initiate the
avalanche. In updating that number and its neighbors, either no number
below $\lambda_{\rm c}$ will be created and that critical avalanche
happens to be empty, or it will produce numbers  below $\lambda_{\rm
c}$ and will progress through more updates until no more numbers below
$\lambda_{\rm c}$ remain at some later time step.  Thereafter, a new
critical avalanche will start at a new, unrelated location. To
describe an individual avalanche, we  can assume that it started
at the origin, and at all times we need only keep track of those
numbers below $\lambda_{\rm c}$. Lattice addresses  appear only as
parameters characterizing exclusively those numbers currently unstable. 
While the number of sites covered by an avalanche $n_{\rm cov}$ can 
grow as much as linear in time $s$, {\em currently\/} unstable sites
at most grow like $s^{1/2}$~\cite{scaling}. Thus, our temporal cut-off
due to memory constraints $N$ behaves like $s_{\rm co}\sim N^2$.

Since our results depend so strongly on results from numerical
simulations, we describe our improved method in great detail.  To
simulate this process we create a list $\cal L$ of currently unstable
numbers, ${\cal L}=\{(\lambda_{\bf r},{\bf r})|\lambda_{\bf
r}<\lambda_{\rm c}\}$. We initialize $\cal L$ with a single entry
$(\lambda_{\bf r}=0,{\bf r}={\bf 0})$. At each update we first remove
the smallest $\lambda_{\bf r}$ in $\cal L$  and any of its neighbors
in the lattice, if those happen to be in $\cal L$.  Then, we draw a
new number for each of those sites, but only sort into  $\cal L$ those
numbers below $\lambda_{\rm c}$ by storing $(\lambda_{\bf r},{\bf
r})$.  Addresses ${\bf r}$  could extend arbitrarily far from the
origin, thus {\em eliminating\/} any spatial cut-off. It is very easy
to keep $\cal L$ compact and sorted according to $\lambda$. In fact,
it is sufficient to sort in a new $\lambda_{\bf r}$ linearly from the
bottom (instead of using a heap, say): the dynamics quickly leads to a
list in which almost all numbers are very densely packed just below
$\lambda_{\rm c}$, and almost every number inserted at the bottom
moves only a few steps through $\cal L$.

Without explicit lattice reference, it is not easy for this  algorithm
to check the minimum's neighbors in the lattice which, if unstable,
would need to be removed from  $\cal L$. To be sure, we would have to
search $\cal L$ {\em at each update\/} to eliminate those neighbors,
which would be unreasonably time consuming. Instead, we utilize a
procedure similar to hash tables~\cite{hash}: When a lattice address
${\bf r}$ with $\lambda_{\bf r}<\lambda_{\rm c}$ is stored in some
entry ${\cal L}_k$, ${\bf r}$ is mapped into an index $i=i({\bf r})$
for a large, sparse array $\cal A$, $|{\cal A}|\gg|{\cal L}|$. ${\cal
A}_i$ in turn stores the index $k$ of ${\cal L}_k$, or is empty
otherwise.  When  activity returns to the neighborhood of ${\bf r}$,
$i({\bf r})$ is calculated and ${\cal A}_i$ can be checked at once to
track an unstable number in ${\cal L}$. It is crucial that $i({\bf
r})$ is unique, of course, but since $|{\cal A}|$ is finite there
exist lattice addresses ${\bf r}\not={\bf r'}$ such that $i({\bf
r})=i({\bf r'})$. Those conflicts are rare between any two {\em
currently\/} unstable sites, if  $|{\cal A}|\gg|{\cal L}|$. They can
be resolved by moving to the next available index $i({\bf r})+I$,
where $I$ is the offset to the nearest free entry in ${\cal A}$ (for
important details, see Ref.~\cite{bsd.pre}).

We have first used this simulated branching process to extrapolate for
the values of $\lambda_{\rm c}$~\cite{bsd.pre} given in
Tab.~\ref{lambdac} by simulating avalanches at various $\lambda$ below
$\lambda_{\rm c}$ (see Ref.~\cite{scaling}).  This extrapolation uses
the domain covered by an avalanche, $n_{\rm cov}$, requiring ${\cal
L}$ to record all currently {\em and\/} previously unstable sites. At
$\lambda_{\rm c}$, $n_{\rm cov}$ can increase as much as linear with
$s$, quickly exhausting memory (see below). Fortunately, for $\lambda$
well below $\lambda_{\rm c}$, avalanche duration and coverage are
quickly cut off.

\begin{table}
\caption{ List of the critical threshold values $\lambda_{\rm c}$, and
the exponents $D$ and $\mu$, for $1\leq d\leq6$. The relative error
for $\lambda_{\rm c}$ is of the order of $10^{-5}$. The errors in the
exponent are about $1\%$.  The values for $d=1$,2 are taken from
Ref.~\protect\cite{scaling}.  }
\begin{tabular}{rlll}
\toprule $d$ & $\lambda_{\rm c}$ & $D$ & $\mu$\\ \colrule 1 & 0.66702
 & 2.43 & 0.41 \\ 2 & 0.328855 & 2.92 & 0.68 \\ 3 & 0.201190 & 3.35 &
 0.905 \\ 4 & 0.142412 & 4 ($\log$-corr.) & 0.99 \\ 5 & 0.110020 & 4 &
 1 \\ 6 & 0.089752 & 4 & 1 \\ \botrule
\end{tabular}
\label{lambdac}
\end{table}

\begin{figure}
\vskip 2.2in \includegraphics{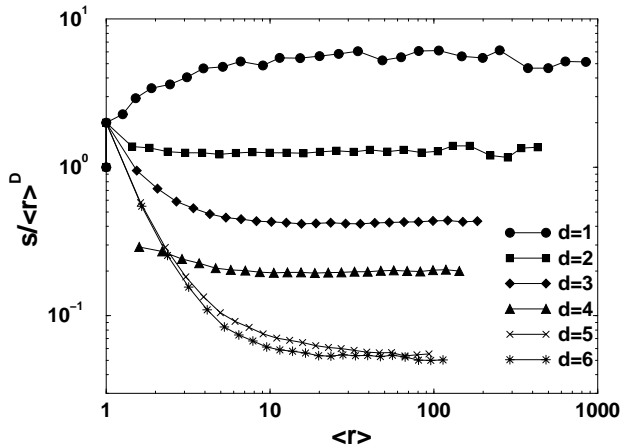}
\caption{ Plot of $s/\langle r\rangle^D$ as a function of $\langle
r\rangle$ for the distribution of avalanche activity in $1\leq
d\leq6$. For $d<4$ we used the fitted values $D_{d=1}=2.43$,
$D_{d=2}=2.92$, and $D_{d=3}=3.35$.  For $d=5$ and 6 we used the
mean-field value $D=4$. For $d=4$ we used $D=4$ as well, but also a
factor of $\ln\langle r\rangle$ as indicated by
Eq.~(\protect\ref{Dscaleq}).  }
\label{plot_D}
\end{figure}

We have run up to $2\times10^6$ critical avalanches in each $d$,
$1\leq d\leq6$, to determine their statistical properties.  We used
$|{\cal L}|_{\rm max}=2^{16}$ and $|{\cal A}|=2^{20}$, easily
sufficient to run avalanches up to $s_{\rm max}=10^8\ll|{\cal L}|_{\rm
max}^2$ update steps. [Avalanches were terminated at $s_{\rm max}$ due
to time constraints and due to the error in $\lambda_{\rm c}$, see
Tab.~\ref{lambdac}.] ${\cal A}$ is already much larger than the
lattice used in Ref.~\cite{highd}, but we point out that even on a
fixed lattice with $N=2^{24}$ sites in $d=6$, the largest distance is
$r_{\rm max}=48$, i.~e. $s_{\rm co}<10^6$. In contrast, our algorithm
has $s_{\rm co}\sim|{\cal L}|_{\rm max}^2\sim5\times10^9$ in any
dimension, yielding significant spatial correlations for $r>100$ even
in $d=6$ (see Fig.~\ref{plot_D}).

In a long critical avalanche, almost all unstable numbers cluster
densely in $\cal L$ with values just below $\lambda_{\rm c}$, which
puts high demands on a random number generator. Typically, in an
ongoing avalanche that has grown to $10^4$ unstable numbers at some
time $s$, $10^3$ of those are placed within
$\Delta\lambda/\lambda_{\rm c}\approx10^{-4}$ just below $\lambda_{\rm
c}$. Using $\sigma=1/2$, those numbers can expect to survive up to
$s\sim(\Delta\lambda/\lambda_{\rm c})^{-1/\sigma}\approx 10^8$ update
steps. To record temporal correlations accurately, it is crucial to
resolve these numbers accurately. This requires numbers that are
sufficiently random for more than $8$ digits! Our data has been
obtained with a sophisticated 64-bit random number generator provided
to us by the authors of Ref.~\cite{RNG}.

First, we discuss the data for the distribution of avalanche activity,
$P(r,s)$, which leads to the avalanche dimension exponent $D$. To find
$P(r,s)$, we record the instances of having activity at a (positive)
distance $r$ relative to the origin at update step $s$. Similar to a
random walk, the moments of the distribution define the exponent $D$
via $\langle r^q\rangle_s\sim s^{q/D}$. In Fig.~\ref{plot_D} we have
plotted the data as $s/\langle r\rangle^D$ vs.  $\langle r\rangle$ for
$d\not=4$.  For $d<4$ we have used the value of our best fit for the
exponent $D$~\cite{scaling,bsd.pre} as given for each dimension in
Tab.~\ref{lambdac}. For $d\geq4$, the fitted value corresponded to the
mean-field result~\cite{BoPa1}, $D=4$.  For $d=4$, we conjecture a
simple asymptotic form for the scaling behavior,
\begin{eqnarray}
s\sim{\langle r\rangle^{4}\over \ln\langle r\rangle}\quad(d=4).
\label{Dscaleq}
\end{eqnarray}
Thus, in Fig.~\ref{plot_D}, unlike for the other dimensions $d$, for
$d=4$ we have plotted $s\ln\langle r\rangle/\langle r\rangle^4$.  For
all dimensions the data levels out horizontally for increasing
$\langle r\rangle$.  Logarithmic factors, as seen in
Eq.~(\ref{Dscaleq}) for $d=4$, are common for scaling behavior at the
upper critical dimension~\cite{logscale}.

\begin{figure}
\vskip 2.2in \includegraphics{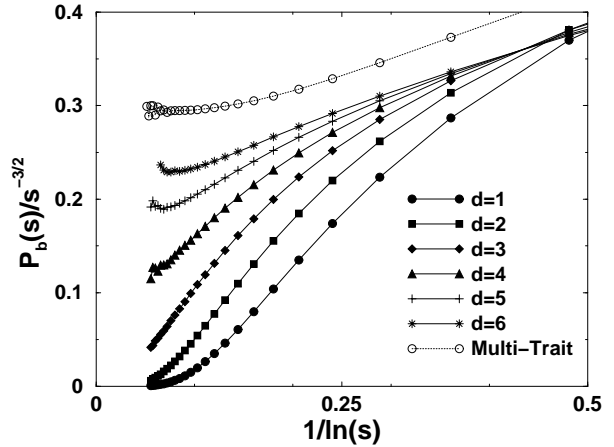}
\caption{Plot of $P_{\rm b}^{\rm all}(s)/s^{-3/2}$ as a function of
$1/\ln(s)$ for backward avalanches for $d=1$ to $6$ and for the
mean-field multi-trait model. For $d<4$ the asymptotic behavior,
$1/\ln(s)\to0$, is characterized by a rapid approach to zero, while
for $d>4$ the ratio does not approach zero. In the marginal case,
$d=4$, $P_{\rm b}^{\rm all}(s)/s^{-3/2}$ appears to vanish as some
power of $1/\ln(s)$. (The multi-trait data has been shifted upward by
$0.1$ for clarity.)}
\label{extra_tau}
\end{figure}

Similar evidence for $d_{\rm c}=4$ is provided by the data for the
backward avalanche exponent $\tau_{\rm b}^{\rm all}$, which is related
to the avalanche distribution exponent through $\tau=3-\tau_{\rm
b}^{\rm all}$~\cite{scaling}. This particular scaling relation does
not rely on avalanches being compact and is valid both above and below
the upper critical dimension.  The data exhibits mean-field behavior,
$\tau_{\rm b}^{\rm all}=3/2(=\tau)$, already for $d=5$ and 6. In
Fig.~\ref{extra_tau} we show the data for the backward-avalanche
distribution reduced by the mean field scaling behavior, $P_{\rm
b}^{\rm all}(s)/s^{-3/2}$ as a function of $1/\ln(s)$, for $d=1$
through $d=6$, and for the multi-trait model~\cite{BoPa1}, which
analytic calculations show exhibits mean field behavior.  The curves
for $d<4$ clearly approach zero rapidly for increasing times $s$ on
this scale, indicating that $\tau_{\rm b}^{\rm all}>3/2$ in those
dimensions. The corresponding data for $d>4$ is clearly bounded away
from zero, comparable to the data from the multi-trait model, where
$\tau_{\rm b}^{\rm all}=3/2$ exactly.  This data also shows that there
are significant corrections to scaling and cut-off effects, even for
the multi-trait model. In this case, simply fitting a power law over
the largest region available from the simulation data can give
misleading results due to systematic curvature on increasing scales.
(Fitted values for $\tau_{\rm b}^{\rm all}$ are discussed in
Ref.~\cite{bsd.pre}.) In $d=4$, the data is again consistent with some
logarithmic correction to scaling behavior.

\begin{figure}
\vskip 2.2in \includegraphics{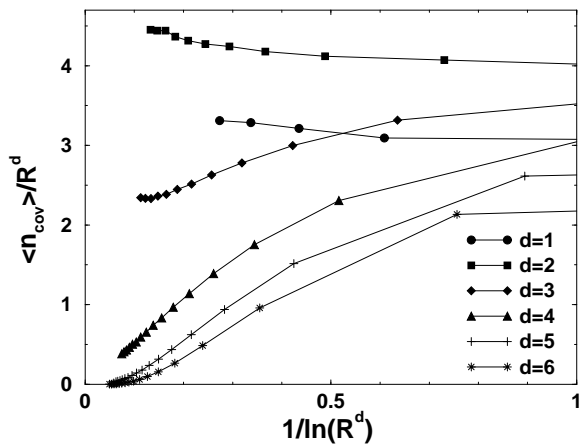}
\caption{ Plot of $\langle n_{\rm cov}\rangle/ R^{d}$ as a function of
$1/\ln(R^d)$ obtained from the coverage $n_{\rm cov}$ and the gyration
radius $R$ of each avalanche in $d=1$ to $6$. Here, the quantity
appears to converge to a finite limit as $1/\ln(R^d)\to0$ for $d<4$,
signaling the compactness of avalanches. In turn, for $d>4$ it rapid
approaches to zero.  In the marginal case, $d=4$, $\langle n_{\rm
cov}\rangle/ R^{d}$ appears to vanish linearly in $1/\ln(R^d)$.  }
\label{extra_df}
\end{figure}

These findings receive strong theoretical support.  The only
assumption in the scaling theory of Ref.~\cite{scaling} is that the
avalanches are compact below the upper critical dimension where each
site that is visited in an avalanche is typically visited many
times. Our numerical results indicate that avalanches indeed remain
compact for $d\leq4$. To test this explicitly, we have determined the
probability distribution $P(n_{\rm cov},R)$ of having a finished
avalanche that has covered a domain $n_{\rm cov}$ with a radius of
gyration $R$. As mentioned above, $n_{\rm cov}$ may grow linear with
$s$, rapidly exhausting memory for ${\cal L}$. Thus, we used $s_{\rm
max} =|{\cal L}|_{\rm max}= 2^{20}\approx10^6$ and $|{\cal
A}|=2^{22}$.  The exponent $d_{cov}$ is defined through the moments of
the distribution $P(n_{\rm cov},R)$ via $\langle n_{\rm
cov}\rangle\sim R^{d_{cov}}$ for large $R$, providing a measure of the
fractal structure of avalanches if $d_{cov}<d$.  In turn, avalanches
are compact if $d_{cov}=d$. Using our numerical results, we plot in
Fig.~\ref{extra_df} the quantity $\langle n_{\rm cov}\rangle/R^{d}$ as
a function of $1/\ln(R^d)$. If avalanches are fractal ($d_{cov}<d$),
we would find that asymptotically this quantity should approach zero
(like $R^{-(d-d_{cov})}$). In fact, we find that this quantity clearly
remains finite for large $R$ for all dimensions $d<4$. For $d>4$, we
find a rapid approach to zero on this logarithmic scale, indicating
fractal behavior. In $d=4$ this quantity appears to vanish as well,
but as a linear function of $1/\ln(R^d)$. Similar to
Eq.~(\ref{Dscaleq}), we find that the numerical simulations indicate
marginal behavior in $d=4$ with logarithmic corrections.  Thus, in
$d=4$ avalanches are marginally compact (a ``fat fractal''), and the
conditions underlying the scaling theory in Ref.~\cite{scaling} are
upheld for all $d\leq 4$, implying $d_{\rm c}=4$.

\begin{figure}
\vskip 2.2in \includegraphics{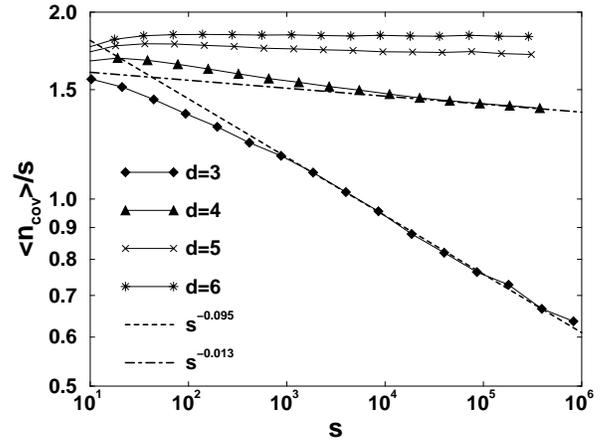}
\caption{ Plot of the coverage, reduced by its mean-field scaling,
$\langle n_{\rm cov}\rangle/s$ as a function of $s$. The data for
$d=5$ and 6 is horizontally level over many decades, consistent with
mean-field behavior. The data for $d=3$ exhibits scaling consistent
with its value of $D$. The deviations from mean-field scaling in $d=4$
are insignificant.  }
\label{plot_mu}
\end{figure}

We demonstrate the consistency of our argument by also measuring the
scaling of the coverage with the lifetime of an avalanche, $\langle
n_{cov}\rangle\!\sim\!s^{\mu}$~\cite{highd}. Since there should be
only one characteristic length for a compact avalanche, we expect that
$\langle r\rangle\sim R$, and thus, $\mu\!=\!d/D$ for $d\leq D$, and
$\mu=1$ in the mean-field limit. In Fig.~\ref{plot_mu}, we plot the
coverage reduced by its mean-field scaling, $\langle
n_{cov}\rangle/s$, as a function of duration $s$. Clearly, the data
for $d\!=\!5$ and 6 is in perfect agreement with mean-field behavior,
while for $d\!=\!4$ the deviation from mean-field behavior are minute,
$\mu_{\rm meas}\!\approx\!0.99$, even without considering the effect
of logarithmic factors. For $d\!=\!3$ we find $\mu\!\approx\!0.905$,
quite consistent with the value predicted from
$\mu\!=\!d/D\!\approx\!0.896$ using our measured value $D\!=\!3.35$ in
$d\!=\!3$ (see Tab.~\ref{lambdac}).

SB acknowledges helpful discussions with M.~J.~Creutz and
T.~T.~Warnock.

\begin{thebibliography}{}
\bibitem{B+S} P.~Bak and K.~Sneppen, Phys.~Rev.~Lett.~{\bf 71},~4083
(1993).
\bibitem{BTW} P. Bak, C. Tang, and K. Wiesenfeld,
Phys. Rev. Lett. {\bf 59}, 381 (1987); Phys. Rev. A. {\bf 38}, 364
(1988).
\bibitem{Bakbook} P.~Bak, {\it How Nature Works: The Science of Self
Organized Criticality,} (Copernicus, New York, 1996).
\bibitem{PNAS} K. Sneppen, P. Bak, H. Flyvbjerg, and M. H. Jensen,
Proc. Nat. Acad. Sci. {\bf 92}, 5209 (1995); P. Bak and M. Paczuski,
Proc. Nat. Acad. Sci. {\bf 92}, 6689 (1995).
\bibitem{Ito} K. Ito, Phys.~Rev.~E {\bf 52}, 3232 (1995).
\bibitem{scaling} M.~Paczuski, S.~Maslov, and P.~Bak, Phys. Rev. E
{\bf 53}, 414 (1996).
\bibitem{EOperc} S. Boettcher, J. Phys. A {\bf 32}, 5201 (1999).
\bibitem{highd} P. De Los Rios, M. Marsili, and M. Vendruscolo,
Phys. Rev. Lett. {\bf 80}, 5746 (1998).
\bibitem{meanfield} H. Flyvbjerg, K. Sneppen, and P. Bak,
Phys. Rev. Lett. {\bf 71}, 4087 (1993); J. de Boer, B. Derrida,
H. Flyvbjerg, A. D. Jackson, and T.  Wettig, Phys. Rev. Lett. {\bf
73}, 906 (1994); J. de Boer, A. D.  Jackson, and T. Wettig,
Phys. Rev. E {\bf 51}, 1059 (1995).
\bibitem{BoPa1} S. Boettcher and M. Paczuski, Phys. Rev. Lett. {\bf
76}, 348 (1996), and Phys. Rev. E {\bf 54} 1082 (1996).
\bibitem{bsd.pre} S. Boettcher, in preparation.
\bibitem{PMB} M. Paczuski, S. Maslov, and P. Bak, Europhys. Lett. {\bf
27}, 97 (1994).
\bibitem{hash} B.~W.~Kernighan and R.~Pike, {\it The Practice of
Programming \/} (Addison-Wesley, Reading, 1999).
\bibitem{RNG} T. T. Warnock, W. W. Wood, and W. Beyer, {\em A new
class of random number generators required for advanced computer
architectures\/}, LA-UR-96-1885, Los Alamos National Laboratory
(1996).
\bibitem{logscale} O. G. Mouritsen, {\em Computer Studies of Phase
Transitions and Critical Phenomena,\/} Sect. 4.2 (Springer, Berlin,
1984).
\end{thebibliography}
\end{document}